\documentclass[prl,twocolumn,superscriptaddress,floatfix,showpacs]{revtex4}

\usepackage[dvips]{graphicx}

\begin{document}

\draft

\title{Calculating Thermodynamics Properties of Quantum Systems by a non-Markovian Monte Carlo Procedure.}

\author{Yanier Crespo}
\affiliation{International School for Advanced Studies (SISSA), and 
Democritos CNR/INFM National Simulation Center, Via Beirut 2-4, I-34014 Trieste, Italy}

\author{Alessandro Laio}
\affiliation{International School for Advanced Studies (SISSA), and
Democritos CNR/INFM National Simulation Center, Via Beirut 2-4, I-34014 Trieste, Italy}

\author{Giuseppe E. Santoro}
\affiliation{International School for Advanced Studies (SISSA), and
Democritos CNR/INFM National Simulation Center, Via Beirut 2-4, I-34014 Trieste, Italy}
\affiliation{International Center for Theoretical Physics (ICTP), I-34014 Trieste, Italy}

\author{Erio Tosatti}
\affiliation{International School for Advanced Studies (SISSA), and
Democritos CNR/INFM National Simulation Center, Via Beirut 2-4, I-34014 Trieste, Italy}
\affiliation{International Center for Theoretical Physics (ICTP), I-34014 Trieste, Italy}

\begin{abstract}
We present a history-dependent Monte Carlo scheme for the efficient calculation of the free-energy 
of quantum systems, inspired by the Wang-Landau sampling and metadynamics method. When embedded in a 
path integral formulation, it is of general applicability to a large variety of Hamiltonians. 
In the two-dimensional quantum Ising model, chosen here for illustration, the accuracy of 
free energy, critical temperature, and specific heat is demonstrated as a function of simulation time, 
and successfully compared with the best available approaches, particularly the Wang-Landau method 
over two different Monte Carlo procedures.
\end{abstract}

% \section{Article}
% \narrowtext
 \maketitle
%\twocolumn[]

%%%%%%%%%!!!!!!!!!!!!!!!!!!Introduction!!!!!!!!!!!!!!!!%%%%%%%%%%%%%%%%%%%%%%%%%%%%%%%%%%%%%%%%%%%%%
%%%%%%%%%  Motivation and problem %%%%%%%%%%%%%%%%%%%%%%%%%%%%%%%%%%%%%%%%%%%%%%%%%%%%%%%%%%%%%%%%%%

Calculating certain thermodynamical quantities, such as the free-energy (FE) or the entropy, by
Monte Carlo (MC) simulation is a notorious difficult problem. 
The difficulty arises because standard MC \cite{metropolis} is devised so as to
generate configurations $X$ distributed according to their Boltzmann weight $P_X=e^{-\beta E_X}/Z$,
where $E_X$ is the energy of the configuration $X$ and $Z=\sum_{X} e^{-\beta E_X}$ the partition function.
This is efficient if we are interested in calculating quantities like the
average energy $\langle E \rangle = \sum_X E_X P_X$, since the configurations generated by 
MC are just 
those 
that contribute significantly to the average. 
Calculating, however, the free-energy, $F=-\beta^{-1} log(Z)$, requires a knowledge of the partition function $Z$
which is not accurately given by the simulation.

A major step forward, in this respect, came with the Wang-Landau (WL) idea \cite{wang-landau}. 
In a nutshell, since:
\begin{equation} \label{Z_classical:eqn}
Z = \sum_X e^{-\beta E_X} = \int \!dE g(E) e^{-\beta E} \;,
\end{equation}
where $g(E)=\sum_{X} \delta(E-E_X)$ is the density of states with energy $E$,
if we devise a MC that generates configurations distributed according to $1/g(E_X)$, 
then we will effectively reconstruct the full {\em histogram} for $g(E)$ in a single simulation.
This allows computing the partition function $Z$, and hence all thermodynamical quantities, 
at any temperature $T=1/k_B\beta$,  where $k_B$ is the Boltzmann constant. 
This is particularly useful if the system can undergo a first-order phase transition.
Indeed, using the WL approach, the system can diffuse over barriers between different local minima 
following pathways that would 
represent,
in normal finite-T MC, ``rare events''.

This discussion applies to classical systems; How should one proceed for a {\em quantum} system? \cite{PIMC+WL}
Consider, to fix 
ideas, the transverse-field quantum Ising model (QIM):
\begin{equation}
\hat H_{\rm QIM} = - J \sum_{\langle ij \rangle}^N \hat\sigma_i^z \hat\sigma_j^z 
         - h\sum_i^N \hat\sigma_i^z - \Gamma \sum_i^N \hat\sigma_i^x \;,
\end{equation}
were $\hat{\sigma}_i^z,\hat{\sigma}_i^x$ are Pauli matrices, $J>0$ is an exchange constant, 
$h$ and $\Gamma$ are respectively the longitudinal and transverse magnetic field, 
and $\langle ij \rangle$ denotes nearest-neighbors on a lattice of $N$ sites. 
The partition sum $Z_{\rm QIM}=\sum_X \langle X|e^{-\beta \hat H}| X\rangle$, where $X=\{\sigma_{i=1\cdots N}\}$
is a configuration of all $N$ spins, involves now a matrix element of $e^{-\beta \hat H}$. 
The first step towards rewriting it in a form similar to Eq.~(\ref{Z_classical:eqn}) 
consists in performing a Suzuki-Trotter decomposition \cite{suzuki}, 
leading to a path-integral expression 
\begin{equation} \label{Z-quantum:eqn}
Z_{\rm QIM} \approx \sum_{\overline{X}} e^{-\beta {\cal A}(\overline{X})} \;.
\end{equation}

Effectively, we have a classical system with an extra time dimension, whose configurations $\overline{X}$, 
over which we sum, are given by $\overline{X}=\{\sigma_{i=1\cdots N; p=1\cdots P}\}$. The extra index 
$p$ labels the $P$ {\em Trotter slices} in the time direction \cite{pbc:note}. 
In the QIM case, the action ${\cal A}$ reads:
\begin{equation}
{\cal A}(\overline{X}) = N\left[ JU_{\overline{X}} + J^{\small \perp}K_{\overline{X}} - h M_{\overline{X}} 
                                - \frac{P}{\beta} \ln{C} \right] \;,
\end{equation}
where $U_{\overline{X}}= -(NP)^{-1}\sum_p \sum_{\langle ij \rangle} \sigma_{i,p} \sigma_{j,p}$ 
is the classical interaction energy per spin, 
$K_{\overline{X}}= -(NP)^{-1} \sum_{i,p} \sigma_{i,p} \sigma_{i,p+1}$ is the quantum ``kinetic energy'' per spin, 
$M_{\overline{X}}= (NP)^{-1} \sum_{i,p} \sigma_{i,p}$ is the magnetization per spin, 
$J^\perp = -(P/2\beta)\rm{ln}\left[ \rm{tanh} \left(\beta\Gamma/P \right)\right] > 0$
is the ferromagnetic coupling between adjacent spins in the time-direction, 
and $C^2 = (1/2)\sinh{\left( 2\beta\Gamma/P \right)}$. 
By introducing a multi-dimensional density of states 
$
g(U,K,M) = \sum_{\overline{X}} \delta(U-U_{\overline{X}}) \delta(K-K_{\overline{X}}) \delta_(M-M_{\overline{X}})
$
we can easily rewrite:
\begin{eqnarray} 
Z_{\rm QIM} &\approx& \int \! dU dK dM \; g(U,K,M) \; e^{-\beta A(U,K,M)} \nonumber \\
             &=& \int \! dU dK dM \; e^{-\beta F(U,K,M)} \;, 
\end{eqnarray} 
where $A(U,K,M)= N\left[ JU + J^{\small \perp}K - h M -(P/\beta)\ln{C} \right]$, 
and $F(U,K,M)$ defines the FE as a function of $(U,K,M)$.
For $h=0$ the relevant coordinates are two, $U$ and $K$. Using the WL idea to reconstruct $Z_{\rm QIM}$
for all values of $\beta$ and $\Gamma$ requires now sampling a {\em two-dimensional} density of states histogram
$g(U,K)$ in terms of which $Z_{\rm QIM}\propto \int \! dU dK g(U,K) e^{-\beta N (J U + J^{\small \perp} K)}$.
This approach is, however, not very efficient (see below). 

A much more convenient (``state-of-the-art'') route is based on the so-called 
{\em stochastic series expansion} (SSE) \cite{SSE,SSE_QIM}, and involves using a WL
approach to reconstruct the coefficients $g(n)={\rm Tr}(-\hat H)^n$ of a high-temperature 
expansion of the partition function $Z=\sum_n (\beta^n/n!) g(n)$ \cite{Troyer_WL}.
The SSE approach is particularly suited to treat quantum spin systems and other lattice quantum
problems, but is in general not straightforward, for instance, for quantum problems on the continuum.

We propose here a new method to effectively calculate the FE of a quantum system.
Our approach is based on a path-integral formulation and can be easily extended to complicated 
off-lattice quantum problems.
The crucial ingredients were borrowed from the WL method and the {\em metadynamics} approach, 
a method which proved useful for exploring the FE landscape of complex classical
systems \cite{metady_rev} as a function of many collective variables (CVs) 
$\mathbf{S}=(S_1,\cdots S_d)$. 

In metadynamics, sampling is enhanced introducing a history-dependent potential $V_G(\mathbf{S},t)$, 
defined as a sum of Gaussians centered along the ``walk'' in CVs-space, that in time ``flattens'' 
the FE histogram as a function of the CVs: $V_G(\mathbf{S},t\to\infty)\sim -F(\mathbf{S})$ {\cite{meta_proof}}. 
This approach has been mainly used  within molecular dynamics. 
During the simulation the system is guided by the action of two forces, 
the thermodynamic one, which move it towards the local FE minimum, and that due to the
history-dependent potential, which 
pushes it away from local minima.

We show here how to integrate 
metadynamics
in a MC procedure, in particular in a path-integral MC (PIMC),
to sample the FE landscape of quantum systems as a function of physically relevant 
CVs. 
Again, we illustrate this approach in the quantum Ising model 
where we
reconstruct the FE as a function of three CVs, the magnetization
$M$, the potential energy $U$ and the kinetic energy $K$.
As we will show, a calculation performed at a single point $(\beta,\Gamma,h)$ in parameter space, 
is sufficient to obtain 
the FE in a whole neighborhood of that point. 
The method is tested by comparing its efficiency against the state-of-the-art WL-SSE method \cite{Troyer_WL}, or a WL over a standard PIMC \cite{PIMC+WL}: 
we prove that our approach is at least as good as the WL-SSE on a lattice
quantum problem, as well as being physically transparent and easily generalizable to different models.

%%%%%%%%%!!!!!!!!!!!!!!Description of method!!!!!!!!!!!%%%%%%%%%%%%%%%%%%%%%%%%%%%%%%%%%%%%%%%%%%%%%
%
%
Given the classical-like path-integral expression for the 
partition function of our quantum model, for instance 
$Z\approx \sum_{\overline{X}} e^{-\beta {\cal A}(\overline{X})}$, see Eq.~(\ref{Z-quantum:eqn}), 
we first 
define a 
small 
number $d$ of CVs $S_l(\overline{X})$, $l=1\dots d$, which appear in the action 
${\cal A}(\overline{X})=A(\mathbf{S}(\overline{X}))$: 
in the QIM case there are $d=3$ 
physically meaningful
CVs, the potential energy $S_1=U$, 
the kinetic energy $S_2=K$ and the magnetization $S_3=M$, in terms of 
which the action 
is $A(\mathbf{S})=N \left[ JU + J^{\small \perp}K - hM -(P/\beta)\ln{C}\right]$.
Next, we perform a Metropolis walk in configuration space $\left\lbrace \overline{X}\right\rbrace $ in which the 
transition probability from $\overline{X}$ to $\overline{X}'$ is modified 
adding to the action  a history-dependent potential $V_{G}({\mathbf{S}}(\overline{X}),t)$:
\begin{equation} \label{metropolis}
{\mathcal{P}}({\overline{X}}\rightarrow{\overline{X}'},t) \equiv \text{min}
\left[1, e^{-\beta\left( \delta A + \delta V_{G}(t) \right) } \right]
\end{equation}
where $\delta A={\cal A}({\overline{X}}')-{\cal A}({\overline{X}})$ is the change in action
and $\delta V_{G}(t)=V_{G}({\mathbf{S}(\overline{X}')},t) - V_{G}({\mathbf{S}(\overline{X})},t)$. 
Whether or not a move is accepted, we update $V_G$ by adding
to it a small localized repulsive potential (a Gaussian in normal 
metadynamics {\cite{metady_rev}}).
Technically, this is best done by grid-discretizing the CVs-space 
and keeping track of $V_{G}({\mathbf{S}}^{(k)},t)$ only at grid points ${\mathbf{S}}^{(k)}$; 
the value of $V_G$ at a generic point $\mathbf{S}(\overline{X})$ is then calculated by
a linear interpolation ${\cal L}$ from the neighboring grid-values:
$V_{G}(\mathbf{S}(\overline{X}),t) = {\cal L}(V_{G}(\mathbf{S}^{(k)},t))$
where ${\cal L}(\dots)$ is the linear interpolation function, and ${\mathbf{S}}^{(k)}$, $k=1\dots 2d$, 
are the points of the grid nearest-neighbors of $\mathbf{S}(\overline{X})$.
In this scheme, the potential $V_G$ is updated 
on the neighboring grid-points ${\mathbf{S}}^{(k)}$ as:
\begin{equation} \label{hystory-dependent}
V_{G}({\mathbf{S}}^{(k)},t+1) = V_{G}({\mathbf{S}}^{(k)},t) + w \prod^d_{l=1} 
\left( \frac{S^{(k)}_l-S_l(\overline{X})}{\Delta S_l} \pm 1 \right) \;,
\end{equation}
where the ($+$) sign is used if $S^{(k)}_l \leq S_l(\overline{X})$ and the ($-$) sign otherwise, 
$\Delta S_l$ is the spacing of the grid in the $S_l$ direction and $w$ is 
a parameter that determines the speed of the FE reconstruction.
Therefore, like in WL, the acceptance changes every time a move is accepted or rejected, 
and the ``walk'' in configuration space is intrinsically non-Markovian (it depends on the history).
At the beginning of the simulation the potential $V_{G}(\mathbf{S}^{(k)},t=0)$, stored on the grid, is set to zero.
Then, as the system moves in configuration space, $V_G$ is updated at each move as in Eq.~(\ref{hystory-dependent}). 
After a sufficient time, $V_{G}$ will approximately compensate the underlying FE profile {\cite{meta_proof}}.
A further improvement can be obtained by taking as estimator of the FE not
just 
a single profile $V_G$, but the arithmetic average of all the profiles between a ``filling'' time $t_{F}$ and
the total simulation time $t_{tot}$:
\begin{equation} \label{ave_fes}
F({\mathbf{S}}) \approx
-\frac{1}{t_{tot}-t_{F}} \int_{t_{F}}^{t_{tot}} \! dt \; V_{G}(\mathbf{S},t) \;.
\end{equation}
This reduces the error of the method, 
which drops fast
to zero for large $t_{tot}-t_{F}$ {\cite{metady_rev}}.

\begin{figure}[!htb]% 
\vspace{6mm}
\centering
\includegraphics[width=3.0in,angle=0]{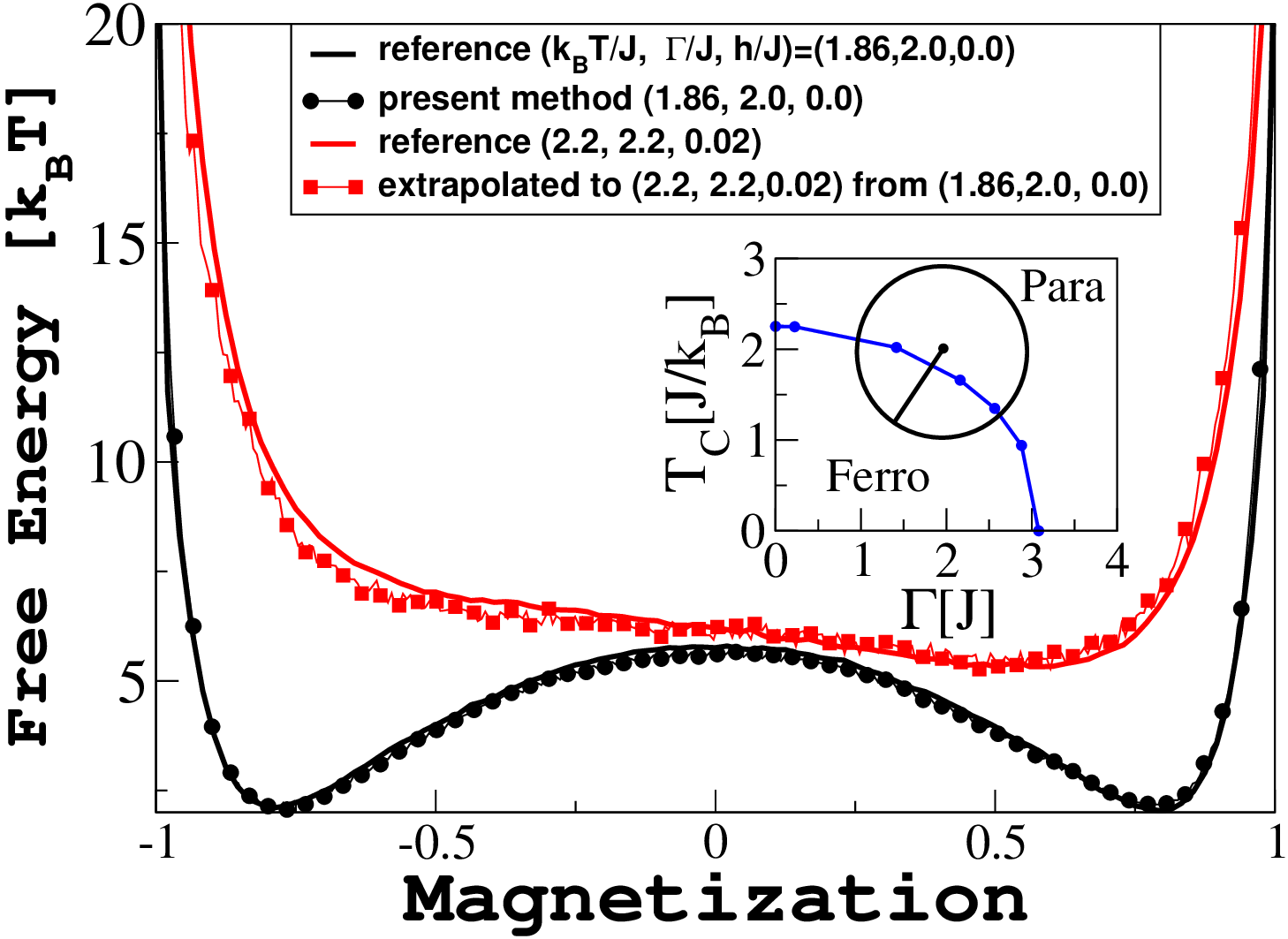}
\caption{ \label{fig:extrap}
Free energy profile for the $8\times8$ QIM, as a function of the magnetization for two different parameter values.
The results at $(k_BT=1.86,\Gamma=2.0,h=0)$ (in units of J) are obtained by first calculating
$F(U,K,M)$, and then performing a logarithmic integration, Eq.~(\ref{LI}), to calculate $F(M)$.
The results at $(k_BT=2.2,\Gamma=2.2,h=0.02$) are instead obtained by extrapolating the previous
$F(U,K,M)$ using Eqs.~(\ref{fe_temp}-\ref{fe_field}), and then integrating to obtain $F(M)$. As a reference 
for the comparison we use the results of an accurate umbrella sampling calculation \cite{umbrella}
(solid line).  
 The inset shows the 
phase diagram of the model and the circle suggests the size of the extrapolation region. 
}
\end{figure}

%If 
When 
$F(U,K,M)$ for a given value of the external parameters $(\beta,\Gamma,h)$ is known, 
one can readily recalculate the new FE profile for a whole neighborhood in parameter space.
The equations for this extrapolation can be written as: 
\begin{widetext}
\begin{eqnarray} \label{fe_temp}
F(U,K,M)_{\beta'} &=& 
\frac{\beta}{\beta'}\left[ F(U,K,M)_{\beta}-N\left( J U +J^\perp_{\beta} K -h M\right)\right] 
+ N\left( J U + J^\perp_{\beta'} K -h M \right) + \frac{NP}{\beta'}\ln\left[\frac{C(\beta)}{C(\beta')}\right] \\
\label{fe_ganma}
F(U,K,M)_{\Gamma'} &=& F(U,K,M)_{\Gamma} + N\left( J^\perp_{\Gamma'} - J^\perp_{\Gamma} \right) K
+ \frac{NP}{\beta}\ln\left[ \frac{C(\Gamma)}{C(\Gamma')}\right] \\
\label{fe_field}
F(U,K,M)_{h'} &=& F(U,K,M)_{h}-N\left(h'-h\right) M \;.
\end{eqnarray}
\end{widetext}
By logarithmic integration of $F(U,K,M)$ with respect to one or more variables we immediately get
the free-energy as a function of a 
reduced 
number of CVs. For instance:
\begin{equation} \label{LI}
F(M) =-\frac{1}{\beta} \ln\left[ \int \! dU dK \; e^{-\beta F(M,U,K)} \right] \;.
\end{equation}
Fig.~ \ref{fig:extrap} shows $F(M)$ for the QIM on a $8\times 8$ lattice ($N=64$ spins), 
with $P=30$ Trotter slices at two different points in parameter space.
The agreement between the reference $F(M)$ and that calculated from $F(U,K,M)$ is  
good, even 
if we extrapolate the $F(U,K,M)$ 
from the ordered to the disordered side (or viceversa) of the phase transition line.
Thus with 
a single 
calculation of $F(U,K,M)$ at a point $(T,\Gamma,h)$ in
parameter space, we can 
get 
reliable 
information for $F(U,K,M)$ in a whole neighborhood of that point (see inset). 
%............................................................................................
% Fig. 1
%............................................................................................
%............................................................................................

In order to test the efficiency of the proposed method we compare it with a SSE-WL
simulation \cite{Troyer_WL,SSE_QIM}, as well as with a direct application of  WL to PIMC 
in which the two-dimensional $g(U,K)$ is calculated.
%
%---------------------------------------------------------------------------------------------------
\begin{figure}[!htb] % fig 3
\vspace{6mm}
\centering
\includegraphics[width=3.2in,angle=0]{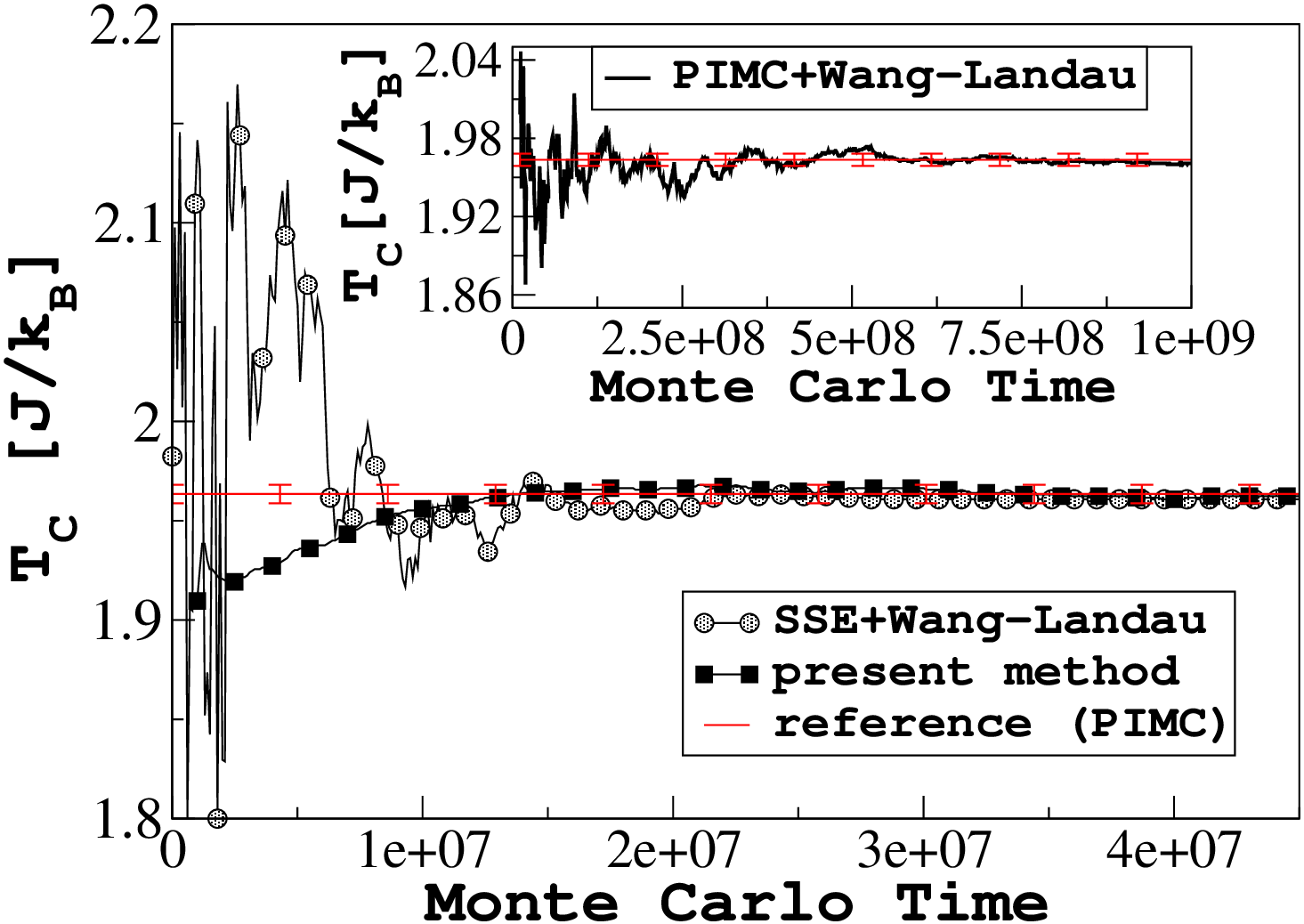}
\caption{\label{fig:cvp8x8}
Critical temperature for a $8\times 8$ QIM, as a function of the MC time,
calculated using three different methods: SSE-WL (solid circles), our method (solid squared)
and the PIMC-WL algorithm (inset). The reference (red line with error bars) is obtained by a long PIMC simulation.}
\end{figure}
%---------------------------------------------------------------------------------------------------
%
For the same system of Fig. \ref{fig:extrap}  we estimate $T_c$ 
(conventionally defined as the temperature at which the specific heat reaches its maximum value)
as a function of the MC time with the three methods.
The results are shown in Fig. \ref{fig:cvp8x8}. 
As a reference, we also computed 
$T_c$ by a very long PIMC calculation (red line with error bars in the Figure).
In the SSE+WL calculation the histogram is considered ``flat'' when for all the values of $n$ the histogram is larger 
than 95 \% of its average {\cite{WL_his_typicall}} (the limit of 80 \% suggested in Ref \cite{wang-landau} leads, for this specific system, 
to systematic errors, data not shown). Instead, for PIMC-WL the  80 \% limit is sufficient to reach convergence.
The specific heat for our method was calculated computing
a $F(U,K)$ at  $k_BT/J=1.8$ and $\Gamma /J=2.0,h/J=0.0$ and extrapolating in temperature according to Eq. (\ref{fe_temp}).
The grid spacing in the $U$ and $K$ directions was of 10 and 1 energy levels respectively. 
However we needed a finer grid spacing of 1 also for $U$, for states with $U < -1.79167$,
in order to avoid systematic errors that generally tend to arise close to the parameter 
boundary values. 

In order to extrapolate the FE in a meaningfull temperature interval $\Delta T \sim \pm J/k_B$ including
the peak of the specific heat, it is necessary to obtain quickly a large maximum value of $V_G \sim 80 k_BT/J$ for
the system considered here. This is accomplished by starting the simulation 
with $w = 8 \cdot 10^{-3}$ decreasing it up to $10^{-4}$ in $2 \cdot 10^6$ MC steps ($t_F$ in Eq. (\ref{ave_fes})), then $w$ is not changed anymore, and the free energy is estimated using  Eq. (\ref{ave_fes}). It is clear from the previous discussion that the optimal ``filling'' protocol is system-dependent.

As shown in Fig. \ref{fig:cvp8x8}, using our approach we can obtain $T_c$ within the PIMC error bar, 
with an efficiency similar to the SSE+WL algorithm.  
The PIMC+WL method is, by comparison, 
an order of magnitude slower (Fig. \ref{fig:cvp8x8}, inset).
Of course, the efficiency of the approach presented here is strongly influenced by the temperature %at which 
where
the reconstruction is performed, %that cannot 
which should not 
be too far from $T_c$ ($\sim 10 \%$ smaller in the example considered here). 
However, $T_c$ can always be estimated {\it approximately, e.g.}, by performing a preliminary calculation on a system of 
smaller size.

%---------------------------------------------------------------------------------------------------
\begin{figure}[!htb]% fig 3
\vspace{6mm}
\centering
\includegraphics[width=3.2in,angle=0]{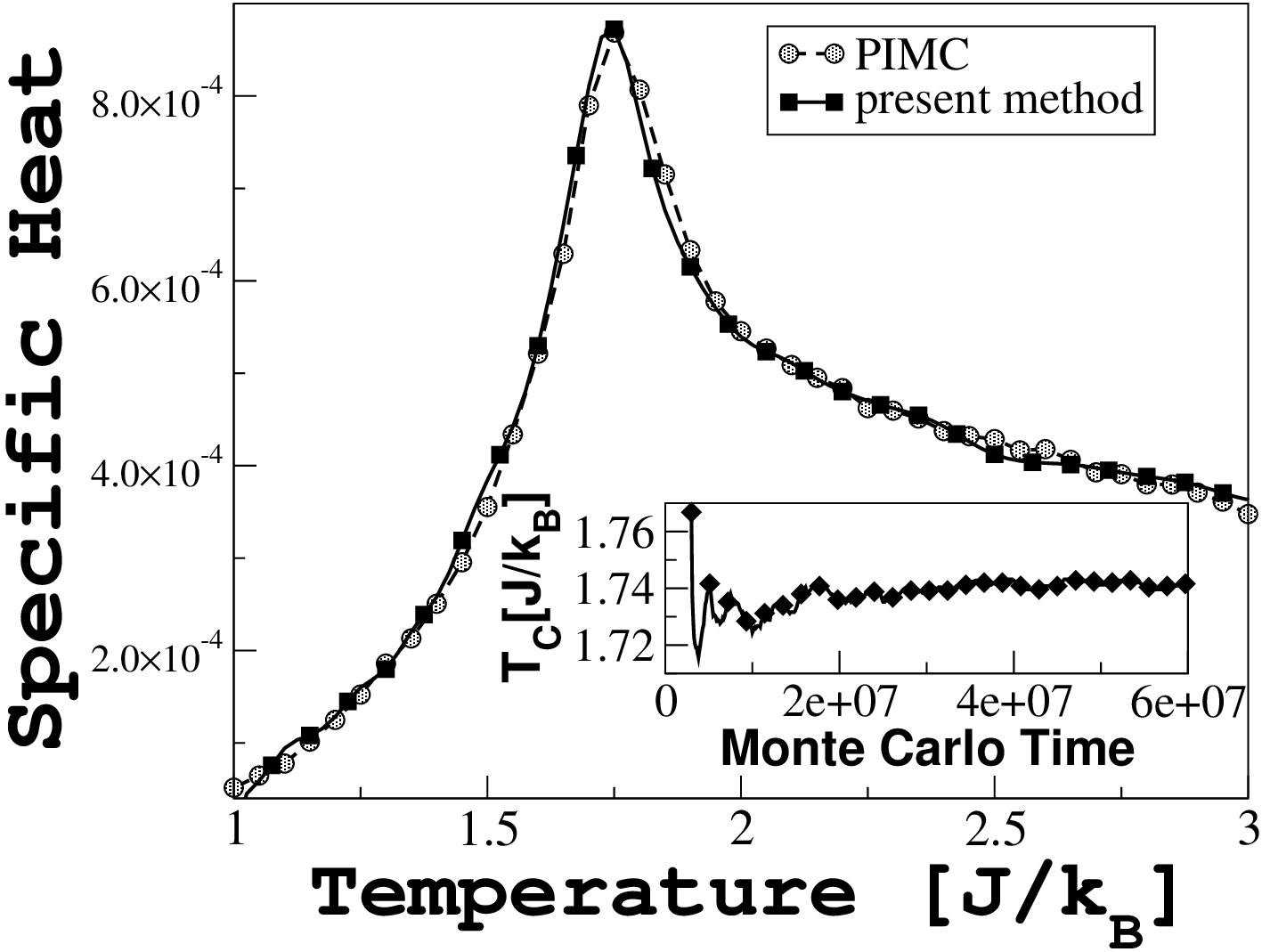}
\caption{\label{fig:quantum_ising}
Specific heat as a function of the temperature for the $32\times32$ QIM using two different methods, 
the PIMC technique (solid circles) and the proposed method (solid squares). 
The inset shows how the estimate of  $T_c$ evolves as a function of the MC time using the present scheme.}
\end{figure}
%---------------------------------------------------------------------------------------------------
 
Fig. \ref{fig:quantum_ising} 
shows 
the specific heat as a function of T
for a larger system, $N=32\times 32$, with $P=100$ Trotter slices, calculated with 
PIMC and 
with the 
present method.  Also in this case we computed $F(U,K)$ and extrapolated
in temperature according to Eq. (\ref{fe_temp}), with a grid spacing of 150 and 10 energy levels 
in $U$ and $K$ directions respectively (
no need to reduce the grid spacing near the parameter boundaries since the 
free energy is very high there). 
In this case $w$ was decreased from $10^{-1}$ to $5 \cdot 10^{-3}$ in $1.1 \cdot 10^6$ MC steps. 
After this time the free energy is estimated using  Eq. (\ref{ave_fes}). As shown 
in Fig. \ref{fig:quantum_ising}, 
our approach reproduces
the specific heat accurately between $1$ and $3 k_BT/J$.
In the inset we show  how $T_c$  converges as a function of the MC time. 
Remarkably, even for this much larger  $32\times32$  
system the convergence of $T_c$ needs roughly the same order of magnitude 
of MC steps of those  needed for the small $8\times 8$ system.

In conclusion, we have introduced an efficient history-dependent Monte Carlo scheme
that allows the accurate calculation of the free energy landscape of quantum systems.
The proposed approach was tested on a two-dimensional quantum Ising model, where we reconstruct 
the free energy as a function of two and three collective variables. This allows 
reproducing the thermodynamic properties in a whole neighborhood of the point 
in parameter space at which the calculation is performed.
The number of MC steps that are necessary to estimate $T_c$ in a 
relatively large system ($32 \times 32\times 100$) is of the same order
 as that required in a small system ($8\times 8\times 30$). The efficiency in estimating $T_c$ is similar 
to that of SSE+WL, the state-of-the-art approach. Based on path-integral MC,
our method can however be directly applied to continuous, off-lattice quantum problems, 
where SSE would be harder to implement.

\begin{acknowledgments}
 This research was partially supported by a MIUR/PRIN contract, and benefited from the environment
 provided by the CNR/ESF/EUROCORES/FANAS/AFRI project.
\end{acknowledgments}

%%%%%%%%%%%%%%%%%%%%%%%%%%% BIBLIO %%%%%%%%%%%%%%%%%%%%%%%%%%%%%%%%%%%%%%%%%%%%%%%%%%%%%%%%%%%%%%%%%55


\begin{references}
\bibitem{pbc:note} Periodic boundary conditions are imposed 
in the time-direction, as dictated by the trace in the quantum partition function.
\end{references}
\end{document}